# К 50-летию дифракционной теории Глаубера[1]

А.А. Архипов

*ФГУП ГНЦ Институт физики высоких энергий*
*Россия, 142280 Московская область, г. Протвино*

**Аннотация**

В данном миниобзоре представлен исторический экскурс в теоретические исследования, истоки которых восходят к работе Глаубера 1955г. по дифракционной теории фундаментальных ядерных процессов, с акцентом на последние достижения в этой области исследований.

## 1 Введение: дифракционная теория Глаубера

Одним из лауреатов Нобелевской премии по физике за 2005г. стал известный американский физик-теоретик Рой Глаубер. Столь престижной международной премией был отмечен его вклад в квантовую теорию оптической когерентности. В России, да и во всем мире, в особенности, в научных кругах, связанных с исследованиями в области физики элементарных частиц и ядерной физики, Глаубер широко известен как основоположник дифракционной теории фундаментальных ядерных процессов. Случайно это или нет, но так получилось, что присуждение Нобелевской премии Р. Глауберу совпало с его восьмидесятилетием и с 50-летием публикации его работы [1], в которой были сформулированы основные положения квантовой теории дифракционного рассеяния быстрых частиц на дейтроне. Нам представляется, что такое совпадение событий (точка тройного совпадения!) располагает к тому, чтобы вспомнить и эту работу Глаубера, сделав некоторую ретроспекцию в сравнении с последними достижениями в этой области исследований.

В научной литературе можно встретить много разных названий для предложенной Глаубером теории, например, эйкональное приближение, глауберовское представление и т.д. Мы, впрочем, как и сам Глаубер, считаем, что лучшим названием теории Глаубера, отражающим наиболее точно ее суть, является «дифракционная теория». Дело в том, что теория Глаубера представляет собой обобщение и распространение известной из физической оптики теории дифракции Фраунгофера на случай ядерных столкновений. При этом основное условие применимости дифракционной теории рассеяния формулируется в виде предположения о том, что импульс налетающей частицы $p$ должен быть настолько большим, чтобы соответствующая этому импульсу длина волны $\lambda = 1/p$ ($\hbar = c = 1$) была много меньше радиуса взаимодействия $r_0$ частицы с нуклоном, входящим в состав ядра,

$$pr_0 = r_0/\lambda \gg 1, \quad r_0 \sim 1.5\,\text{fm}. \tag{1}$$

---

[1] Расширенный текст доклада, заявленного для представления на юбилейной сессии отделения ядерной физики РАН, состоявшейся в ИТЭФ 5-9 декабря 2005г.

Очевидно, что в этом случае в амплитуду упругого рассеяния быстрой частицы дает вклад большое число парциальных волн $l \sim pr_0 \gg 1$, вследствие чего угловое распределение в упругом рассеянии при высоких энергиях сосредоточено в ярко выраженном узком конусе с характерным дифракционным пиком рассеяния вперед.

Упругое рассеяние быстрых частиц на ядрах характеризуется рядом особенностей, которые могут быть легко поняты, поскольку они физически чрезвычайно прозрачны. Поскольку ядра представляют собой слабосвязанные системы нуклонов (энергия связи на нуклон не превосходит нескольких MeV), средние значения импульсов нуклонов в системе покоя ядра невелики, порядка $1/R$, где $R$ среднее межнуклонное расстояние, сравнимое с размером ядра. Налетающая на ядро частица с большой энергией в каждом отдельном столкновении с нуклонами ядра может передать нуклону лишь малую долю своего импульса и энергии. Если $q = (p - p')$ – переданный импульс нуклону ядра ($p$ и $p'$ – импульсы налетающей частицы до и после столкновения с нуклоном ядра), то

$$|\mathbf{q}| \lesssim \frac{1}{R}, \tag{2}$$

поскольку нуклон, испытавший при столкновении с налетающей частицей импульс отдачи больший, чем $1/R$, будет выбит из ядра, ядро развалится, а такие процессы с развалом ядра не могут быть отнесены к процессам упругого рассеяния. Конечно, всегда имеются и такие события, когда ядру как целому передается большой импульс и ядро при этом не разваливается, однако вероятность таких событий очень мала. Основную долю составляют такие события, когда частица высокой энергии, проходя через ядро, лишь слегка отклоняется от своего первоначального направления движения и покидает ядро, не вызвав существенной перестройки последнего. При столкновении быстрой частицы с каким-либо из нуклонов ядра, все остальные нуклоны ядра выступают как бы в роли наблюдателей, не участвующих во взаимодействии.

Предположим теперь, что высокоэнергичная частица, проходя через ядро, рассеивается не более одного раза, т.е. сталкивается только с одним из нуклонов ядра. Поскольку невозможно указать с каким из нуклонов ядра произошло столкновение, то амплитуда упругого рассеяния на ядре в этом случае будет равна когерентной сумме амплитуд рассеяния на каждом из нуклонов. Тот факт, что налетающая частица рассеивается не на свободном нуклоне, а на нуклоне, находящемся в связанном состоянии в ядре можно учесть введением формфактора ядра. Если плотность ядерной материи в ядре обозначить через $\rho(\mathbf{r})$ и нормироать ее условием $\int \rho(\mathbf{r}) d\mathbf{r} = 1$, то формфактор ядра определяется как

$$\Phi(\mathbf{q}) = \int e^{i\mathbf{q}\mathbf{r}} \rho(\mathbf{r}) d\mathbf{r}. \tag{3}$$

Обозначив амплитуду упругого рассеяния на одном нуклоне с переданным импульсом $\mathbf{q}$ через $f_{hN}(E, \mathbf{q})$ ($E$ – энергия налетающей частицы в лабораторной системе), для амплитуды упругого рассеяния на ядре $f_{hA}(E, \mathbf{q})$ в приближении однократного рассеяния получаем

$$f_{hA}(E, \mathbf{q}) = A f_{hN}(E, \mathbf{q}) \Phi(\mathbf{q}). \tag{4}$$

Отсюда для полного сечения рассеяния на ядре и углового распределения следует

$$\sigma_{hA}^{tot}(E) = A \sigma_{hN}^{tot}(E), \qquad \frac{d\sigma_{hA}}{d\Omega} = A^2 \frac{d\sigma_{hN}}{d\Omega} \Phi^2(\mathbf{q}). \tag{5}$$

Присутствие формфактора ядра в угловом распределении приводит к тому, что интенсивность процесса упругого рассеяния быстрых частиц на ядре сосредоточена в еще более узком конусе, угловые размеры которого теперь определяются отношением длины волны падающей частицы к размеру ядра

$$\theta \lesssim \frac{1}{pR} < \frac{1}{pr_0}, \quad r_0 < R. \tag{6}$$

Из формул (5) также следует, что дифференциальное сечение рассеяния на ядрах вперед квадратично зависит от числа нуклонов в ядре и растет как $A^2$. Разумеется, все перечисленные выше выводы основаны на предположении об однократном взаимодействии налетающей частицы с нуклонами ядра и справедливы только в этом приближении.

В экспериментальных исследованиях по измерению сечений рассеяния на дейтроне с пучками протонов, нейтронов и пионов при импульсе налетающих частиц порядка $1 \div 1.5$ GeV ($\lambda \sim 0.1 \div 0.2 \, \text{fm}$), выполненных в начале пятидесятых годов, было обнаружено, что полные сечения рассеяния на дейтроне во всех случаях меньше, чем сумма полных сечений рассеяния на свободных нуклонах. Результаты измерений можно было выразить формулой

$$\sigma_{hd} = \sigma_{hp} + \sigma_{hn} - \delta\sigma_{hd}, \tag{7}$$

где $\sigma_{hd}$, $\sigma_{hp}$, $\sigma_{hn}$ – полные сечения рассеяния налетающей частицы $h$ ($h = p, n, \pi^\pm$) на дейтроне, протоне и нейтроне, величина $\delta\sigma_{hd}$ называется дефектом полного сечения рассеяния на дейтроне. Величина дефекта $\delta\sigma_{hd}$ для пучков нейтронов и протонов составляла $\delta\sigma_{Nd} \sim 6 \div 9 \, \text{mb}$, для пучков пионов – немного меньше $\delta\sigma_{\pi^\pm d} \sim 4 \div 6 \, \text{mb}$ (см. [1] и имеющиеся там ссылки на соответствующие экспериментальные исследования). Данные экспериментальные результаты, собственно, и подвигли Глаубера к теоретическим исследованиям, в которых была предпринята попытка найти объяснение обнаруженного эффекта. В рамках дифракционной теории Глаубер получил изящное выражение для дефекта полного сечения рассеяния на дейтроне, которое имеет вид

$$\delta\sigma_{hd} = \delta\sigma_G = \frac{\sigma_{hp} \cdot \sigma_{hn}}{4\pi} <\frac{1}{r^2}>_d \tag{8}$$

и носит название глауберовской поправки. В формуле (8) $<r^{-2}>_d$ обозначает среднее значение обратного квадрата межнуклонного расстояния в дейтроне.

Глаубер нашел привлекательную физическую интерпретацию найденной поправке, показав, что она связана с такими конфигурациями в дейтроне, когда один из нуклонов находится в области тени другого, и описывает эффект «затмения», хорошо известный из данных астрономических наблюдений по уменьшению светимости бинарных звезд во время затмения. По этой причине эту поправку часто называют теневой поправкой или эффектом экранировки. Кроме того, необходимо отметить и тот замечательный факт, что формулу (8) можно получить, исходя из чрезвычайно простых, почти полуклассических соображений, представленных Глаубером во введении к своей работе [1].

В основе формализма, предложенного Глаубером, лежит также чрезвычайно простая и физически прозрачная идея. Суть ее состоит в том, что весь эффект взаимодействия быстрой частицы сводится к изменению фазы падающей волны, которую в свою очередь можно выразить с помощью известной из оптики формулы для эйконала. Глауберовское представление для амплитуды рассеяния высокоэнергичной частицы записывается в виде двумерного интеграла Фурье

$$f_\Gamma(E; \mathbf{q}) = \frac{i|\mathbf{p}|}{2\pi} \int d^2\mathbf{b}\, e^{i\mathbf{q}\mathbf{b}} \Gamma(E; \mathbf{b}), \qquad (9)$$

где $\Gamma(E; \mathbf{b})$ называется функцией профиля, $\mathbf{q} = \mathbf{p} - \mathbf{p}'$ – переданный импульс, $\mathbf{b}$ – двумерный вектор прицельного параметра, лежащий в плоскости, ортогональной к импульсу $\mathbf{p}$ падающей волны. Ясно, что функцию профиля можно выразить через амплитуду рассеяния с помощью обратного преобразования Фурье

$$\Gamma(E; \mathbf{b}) = \frac{1}{2\pi i|\mathbf{p}|} \int d^2\mathbf{q}\, e^{-i\mathbf{q}\mathbf{b}} f_\Gamma(E; \mathbf{q}), \qquad (10)$$

где интегрирование проводится также в двумерной плоскости, ортогональной к импульсу $\mathbf{p}$. Функция профиля $\Gamma(E; \mathbf{b})$ связана с комплексной функцией эйконала $\chi(E; \mathbf{b})$ формулой

$$\Gamma(E; \mathbf{b}) = 1 - e^{i\chi(E; \mathbf{b})}. \qquad (11)$$

Соотношение (11), связывающее функцию профиля с эйконалом, является основополагающим в подходе Глаубера. Важность этого соотношения обусловлена тем, что оно обеспечивает непротиворечивость дифракционной теории с фундаментальным требованием квантовой теории – унитарностью.

Следующее важное предположение дифракционной теории, навеянное также аналогией с оптикой, гласит, что весь эффект взаимодействия при прохождении падающей волны сквозь ядро сводится к тому, что фазы рассеяния на отдельных нуклонах просто складываются. Например, в случае рассеяния на дейтроне полный сдвиг фазы определяется выражением

$$\chi_d(E; \mathbf{b}, \mathbf{s}_p, \mathbf{s}_n) = \chi_p(E; \mathbf{b} - \mathbf{s}_p) + \chi_n(E; \mathbf{b} - \mathbf{s}_n), \qquad (12)$$

где $\mathbf{s}_p, \mathbf{s}_n$ – проекции координат нуклонов $\mathbf{r}_p, \mathbf{r}_n$ на двумерную плоскость, ортогональную импульсу $\mathbf{p}$ налетающей частицы. С помощью формул (11,12) функцию профиля для дейтрона можно выразить через функции профиля для отдельных нуклонов

$$\Gamma_d(E; \mathbf{b}, \mathbf{s}_p, \mathbf{s}_n) = 1 - e^{i\chi_d(E; \mathbf{b}, \mathbf{s}_p, \mathbf{s}_n)} = 1 - e^{i\chi_p(E; \mathbf{b} - \mathbf{s}_p)} e^{i\chi_n(E; \mathbf{b} - \mathbf{s}_n)}$$

$$= 1 - [1 - \Gamma_p(E; \mathbf{b} - \mathbf{s}_p)][1 - \Gamma_n(E; \mathbf{b} - \mathbf{s}_n)]. \qquad (13)$$

Раскрывая во второй строке равенства (13) произведение, окончательно для функции профиля дейтрона получаем

$$\Gamma_d(E; \mathbf{b}, \mathbf{s}_p, \mathbf{s}_n) = \Gamma_p(E; \mathbf{b} - \mathbf{s}_p) + \Gamma_n(E; \mathbf{b} - \mathbf{s}_n) - \Gamma_p(E; \mathbf{b} - \mathbf{s_p}) \Gamma_n(E; \mathbf{b} - \mathbf{s}_n). \qquad (14)$$

Теперь, чтобы воспользоваться формулой (9) для вычисления амплитуды упругого рассеяния на дейтроне, необходимо функцию профиля дейтрона, заданную соотношением (14), усреднить по основному состоянию дейтрона. Таким образом,

$$f_d(E; \mathbf{q}) = \frac{i|\mathbf{p}|}{2\pi} \int d^2\mathbf{b}\, e^{i\mathbf{q}\mathbf{b}} <d|\Gamma_d(E; \mathbf{b}, \mathbf{s}_p, \mathbf{s}_n)|d> = \frac{i|\mathbf{p}|}{2\pi} \int d^2\mathbf{b}\, e^{i\mathbf{q}\mathbf{b}} \bar{\Gamma}_d(E; \mathbf{b}), \qquad (15)$$

где
$$\bar{\Gamma}_d(E; \mathbf{b}) \equiv <d|\Gamma_d(E; \mathbf{b}, \mathbf{s}_p, \mathbf{s}_n)|d>. \tag{16}$$

Введем формфактор
$$\Phi_d(\mathbf{q}) = \int e^{i\mathbf{q}\mathbf{r}} |\varphi_d(\mathbf{r})|^2 d\mathbf{r}, \tag{17}$$

где $\varphi_d(\mathbf{r})$ – волновая функция основного состояния дейтрона. Тогда, с помощью обратного преобразования Фурье (10), выражающего функцию профиля через амплитуду рассеяния, используя формулу (14) для функции профиля дейтрона, после несложных преобразований амплитуду упругого рассеяния на дейтроне можно представить в виде [1]

$$f_d(E; \mathbf{q}) = f_p(E; \mathbf{q})\Phi_d(\frac{1}{2}\mathbf{q}) + f_n(E; \mathbf{q})\Phi_d(\frac{1}{2}\mathbf{q}) +$$
$$+ \frac{i}{2\pi|\mathbf{p}|} \int d^2\mathbf{q}' \Phi_d(\mathbf{q}') f_p(E; \frac{1}{2}\mathbf{q} + \mathbf{q}') f_n(E; \frac{1}{2}\mathbf{q} - \mathbf{q}'). \tag{18}$$

Формула (18) замечательна тем, что она выражает амплитуду упругого рассеяния на дейтроне через величины, свойства которых хорошо известны непосредственно из эксперимента. К таковым относятся амплитуды рассеяния на нуклонах и формфактор дейтрона. Каждое слагаемое в формуле (18) имеет ясную физическую интерпретацию: первые два слагаемых соответствуют процессам однократного взаимодействия налетающей частицы с нуклонами дейтрона, а третье слагаемое описывает вклад в амплитуду рассеяния на дейтроне процессов двукратного взаимодействия или процессов упругого перерассеяния налетающей частицы на нуклонах дейтрона. Если учесть, что формфактор дейтрона $\Phi_d(\mathbf{q})$ является быстро падающей функцией от $\mathbf{q}^2$ и существенно отличен от нуля лишь вблизи $\mathbf{q}^2 = 0$, то из формулы (18) можно получить более простое выражение для амплитуды упругого рассеяния на дейтроне

$$f_d(E; \mathbf{q}) = f_p(E; \mathbf{q})\Phi_d(\frac{1}{2}\mathbf{q}) + f_n(E; \mathbf{q})\Phi_d(\frac{1}{2}\mathbf{q}) + \frac{i}{|\mathbf{p}|R_d^2} f_p(E; \frac{1}{2}\mathbf{q}) f_n(E; \frac{1}{2}\mathbf{q}) \tag{19}$$

где, по определению, введено обозначение
$$\frac{2\pi}{R_d^2} \stackrel{def}{=} \int d^2\mathbf{q} \, \Phi_d(\mathbf{q}). \tag{20}$$

Введенное обозначение обосновывается тем, что справедливость равенства (20) проверяется интегрированием в квадратурах для формфактора дейтрона $\Phi_d(\mathbf{q}) = \exp(-R_d^2 \mathbf{q}^2/2)$, отвечающего такому гауссову распределению для плотности ядерной материи в дейтроне $\rho_d(\mathbf{r})$, для которого $<1/r^2>_d = 1/R_d^2$. Выражение (19) для амплитуды упругого рассеяния на дейтроне возникает, если интеграл в третьем слагаемом формулы (18) оценить, вынося медленно меняющиеся по сравнению с формфактором дейтрона амплитуды упругого рассеяния на нуклонах за знак интеграла в точке $\mathbf{q}' = 0$.

Используя оптическую теорему, которая связывает полное сечение взаимодействия с мнимой частью амплитуды упругого рассеяния вперед

$$\sigma^{tot}(E) = \frac{4\pi}{|\mathbf{p}|} \Im f(E; \mathbf{q} = 0), \tag{21}$$

из формулы (19) получаем

$$\sigma_d^{tot}(E) = \sigma_p^{tot}(E) + \sigma_n^{tot}(E) - \delta\sigma_d(E), \qquad (22)$$

где

$$\delta\sigma_d(E) = -\frac{4\pi}{|\mathbf{p}|^2 R_d^2}\Re[f_p(E;\mathbf{0})f_n(E;\mathbf{0})]. \qquad (23)$$

Если еще предположить, что амплитуды упругого рассеяния вперед на нуклонах при высоких энергиях становятся чисто мнимыми, то из формулы (23) с учетом оптической теоремы (21) получаем выражение для глауберовской поправки

$$\delta\sigma_d(E) = \delta\sigma_G(E) = \frac{\sigma_p^{tot}(E)\sigma_n^{tot}(E)}{4\pi R_d^2}. \qquad (24)$$

Для проверки теории в распоряжении Глаубера имелись экспериментальные данные по полным нуклон-нуклонным и нуклон-дейтронным сечениям при импульсе налетающих протонов и нейтронов 1.4 GeV. Данные были таковы

$$\sigma_{pp}^{tot} = \sigma_{nn}^{tot} = 48\,\text{mb}, \quad \sigma_{pn}^{tot} = \sigma_{np}^{tot} = 42\,\text{mb},$$

и

$$\sigma_{pd}^{tot} = 81\,\text{mb}, \quad \sigma_{nd}^{tot} = 84\,\text{mb}.$$

Из этих данных следовало, что дефект полного сечения рассеяния на дейтроне составлял

$$\delta\sigma_{pd}^{tot} = 9\,\text{mb}, \quad \delta\sigma_{nd}^{tot} = 6\,\text{mb}. \qquad (25)$$

Используя для радиуса дейтрона величину $R_d \simeq 2\,\text{fm}$, из формулы (24) для величины глауберовской поправки находим

$$\delta\sigma_{pd}^{tot} = \delta\sigma_{nd}^{tot} \simeq 4\,\text{mb}.$$

Видно, что глауберовская поправка (24), давая правильную по порядку величину, приводит все-таки к заниженному значению для величины дефекта полного сечения рассеяния на дейтроне.

Возвращаясь к выводу формулы (24), можно сделать ряд уточнений. Во-первых, можно отказаться от предположения о том, что амплитуды упругого рассеяния вперед на нуклонах чисто мнимы, и учесть реальную часть этих амплитуд. Кроме того, для интеграла в третьем слагаемом формулы (18) можно предъявить более точное выражение, чем сделанная выше оценка, если учесть, что дифференциальное сечение упругого рассеяния на нуклонах при высоких энергиях, как уже отмечалось выше, имеет ярко выраженный дифракционный пик вперед, который можно смоделировать путем представления мнимой части амплитуды упругого рассеяния в виде

$$\Im f_N(E;\mathbf{q}) = \frac{|\mathbf{p}|}{4\pi}\sigma_N^{tot}(E)\exp(-\frac{B_N(E)}{2}\mathbf{q}^2), \quad N = p, n, \qquad (26)$$

где $B_N(E)$ характеризует наклон дифракционного пика в упругом рассеянии на нуклонах. В результате указанных уточнений выражение (24) для глауберовской поправки видоизменяется

$$\delta\sigma_d(E) = \frac{\sigma_p^{tot}(E)\sigma_n^{tot}(E)}{4\pi[R_d^2 + B_p(E) + B_n(E)]}\Big(1 - \rho_p(E)\rho_n(E)\Big), \qquad (27)$$

где
$$\rho_N(E) = \Re f_N(E;\mathbf{0})/\Im f_N(E;\mathbf{0}), \quad N = p, n.$$

Хотя величины $\rho_N(E)$ и $B_N(E)$ ($N = p, n$) при энергиях налетающих частиц 1÷1.5 GeV, обсуждаемых в работе Глаубера, относительно малы ($\rho_N(E) << 1$, $B_N(E) << R_d^2$) и их введение в глауберовскую поправку не приводит к ее существенному численному изменению, тем не менее, необходимо отметить, что оба вышеуказанных уточнения работают на увеличение расхождения теории с экспериментальными измерениями. В качестве одного из решений проблемы устранения расхождения теории с экспериментом, Глаубер видел в построении модели таких межнуклонных сил в дейтроне, которые создавали бы повышенную концентрацию ядерной материи вблизи центра, придавая дейтрону более компактную форму. Кандидатом на такую модель, по мнению Глаубера, мог бы служить потенциал Хюльтена $V(r) \sim \lambda(e^{\lambda r} - 1)^{-1}$ ($\lambda \sim 1.21\,\text{fm}^{-1}$), который замечателен тем, что передает основные черты мезонной теории ядерных сил; он так же, как и потенциал Юкавы, сингулярен в начале как $1/r$ и экспоненциально убывает на бесконечности. Глауберовская поправка, вычисленная в модели межнуклонных сил с потенциалом Хюльтена оказалась равной $\delta\sigma_d = 7.2$ mb, что находится примерно посредине между измеренными значениями (25). Однако достигнутое, как будто бы хорошее согласие с экспериментальными измерениями на самом деле сталкивается с серьезными противоречиями. Дело в том, что требование использовать компактную дейтронную модель с более плотной упаковкой нуклонов в дейтроне не совместимо с основными посылами теории, в которой глауберовская поправка возникает за счет механизма упругого перерассеяния налетающей быстрой частицы на пространственно разделенных нуклонах. Только при таких условиях, когда нуклоны в дейтроне пространственно разделены и находятся достаточно далеко друг от друга, представление полного сдвига фазы рассеяния на дейтроне в виде аддитивной суммы сдвигов фаз рассеяния на отдельных нуклонах, положенное в основу глауберовского подхода, можно считать приемлемым. Однако, очевидно, что в дейтроне динамически возможны и такие конфигурации, когда нуклоны находятся вблизи друг от друга. Рассмотрение и учет таких конфигураций, конечно же, требует более детального исследования динамики процессов рассеяния на дейтроне.

Мы намеренно столь подробно коснулись оптической идеи, положеннной в основу дифракционной теории рассеяния, и предложенного Глаубером метода, обслуживающего эту идею, отчасти еще и потому, дабы предоставить возможность познакомиться с ними как можно более широкой аудитории русскоязычных читателей, для которых доступ к оригинальной работе Глаубера [1] может быть затруднителен.

## 2 Дальнейшее развитие и уточнение дифракционной теории Глаубера

Физическая природа расхождения описанной выше глауберовской картины рассеяния на дейтроне с экспериментом была также понятна еще и по следующим соображениям. Хорошо известно, что в релятивистской квантовой теории при больших энергиях возможны процессы рождения частиц даже при очень маленьких переданных импульсах. Рассмотрим для определенности процесс $h + N \to X + N$ рождения адронной системы $X$ в результате взаимодействия налетающего адрона $h$ высокой энергии с нуклоном $N$. Адронная система $X$ может представлять собой либо возбужденное состояние налетающего адрона $h$ либо систему (пучок, струю) частиц с квантовыми числами адрона $h$, возникших в результате взаимодействия высокоэнергичного адрона $h$ с нуклоном. Такие процессы имеют специальное название – они называются процессами дифракционной диссоциации. Пусть импульс налетающего адрона $p_h$, а импульс, переданный нуклону, обозначим через $q$, тогда для инвариантной массы $M_X$ рожденной системы получаем

$$M_X^2 = (p_h - q)^2 = m_h^2 + q^2 - 2E(\mathbf{p}_h)q^0 + 2|\mathbf{p}_h|q_\shortparallel, \qquad (28)$$

где $E(\mathbf{p}_h) = \sqrt{\mathbf{p}_h^2 + m_h^2}$, $q_\shortparallel$ – проекция переданного нуклону импульса $\mathbf{q}$ на направление импульса налетающего адрона. Поскольку в ядре $q^0 \sim \mathbf{q}^2/2m_N$, а в силу (2) $|\mathbf{q}| \lesssim 1/R$, то из равенства (28) следует оценка на величину энергии налетающего адрона, которая необходима для рождения адронной системы с массой $M_X$

$$E(\mathbf{p}_h) \simeq |\mathbf{p}_h| \simeq \frac{M_X^2 - m_h^2}{2q_\shortparallel} \gtrsim (M_X^2 - m_h^2)\frac{R}{2}. \qquad (29)$$

С другой стороны, если учесть, что $s = (p_h + p_N)^2 = m_h^2 + m_N^2 + 2m_N E(\mathbf{p}_h)$, то из равенства (28) следует оценка на величину массы адронной системы $X$, рождение которой возможно без развала ядра

$$M_X^2 \lesssim m_h^2 + \epsilon s, \quad \epsilon \sim \frac{1}{m_N R}. \qquad (30)$$

Если налетающий адрон является $\pi$-мезоном, тогда (полагая в оценке (29) $R \sim 2\,\mathrm{fm}$) находим, что для рождения системы с массой $M_X = 2m_\pi \div 3m_\pi$ импульс налетающего $\pi$-мезона должен быть $|\mathbf{p}_\pi| \gtrsim 300 \div 800\,\mathrm{MeV}$. Из той же оценки (29) следует, что если налетающим адроном является нуклон, то рождение системы с массой $M_X = m_N + m_\pi$ возможно при импульсе налетающего нуклона $|\mathbf{p}_N| \gtrsim 1.4\,\mathrm{GeV}$. Эти достаточно простые рассуждения с количественными оценками показывают, что с увеличением энергии кроме процессов упругого перерассеяния на нуклонах дейтрона (ядра) необходимо также учитывать процессы дифракционной диссоциации в результате неупругого взаимодействия налетающего адрона с нуклонами ядра. Физическая картина, которая при этом возникает, чрезвычайно проста и наглядна. Адрон, налетающий на дейтрон, при неупругом взаимодействии с нуклоном диссоциирует в адронную систему $X$, которая затем поглощается другим нуклоном с переходом этой системы в состояние исходного налетающего адрона так, что ядро в целом не претерпевает существенной перестройки за время, проведенное налетающим адроном внутри ядра. Совершенно ясно, что учет таких процессов приведет к увеличению дефекта сечения и тем самым к улучшению согласия теории с экспериментом.

Дальнейшее развитие и уточнение дифракционной теории Глаубера было связано в основном с разного рода попытками учесть вклад процессов дифракционной диссоциации в рассеяние быстрых частиц на ядрах [2, 3, 4, 5, 6, 7]. Надо заметить, что точный расчет основных характеристик процессов дифракционной диссоциации представляет собой очень сложную, нерешенную до настоящего времени задачу. Поэтому бо́льшая часть исследований в этой области проводилась в рамках представлений, основанных на реджевской феноменологии. Если предположить, как это обычно делается в исследованиях подобного толка, что асимптотика процессов упругого рассеяния и дифракционной диссоциации при высоких энергиях определяется полюсом Померанчука и связанными с ним ветвлениями, то оказывается возможным получить достаточно простое выражение для дефекта полного сечения рассеяния на дейтроне. Такое выражение было получено Грибовым [4] и имеет вид

$$\delta\sigma_d^\Gamma = 2\int d\mathbf{q}^2 \rho(4\mathbf{q}^2)\frac{d\sigma_N}{d\mathbf{q}^2}, \qquad (31)$$

где $\rho(\mathbf{q}^2)$ – электромагнитный (зарядовый) формфактор дейтрона, $d\sigma_N/d\mathbf{q}^2$ – сумма сечений всех процессов, которые могут происходить при взаимодействии налетающего адрона с нуклоном при заданном квадрате $\mathbf{q}^2$ трехмерного переданного импульса нуклону. Реально, величина $d\sigma_N/d\mathbf{q}^2$, стоящая в правой части формулы (31), представляет собой сечение одночастичной инклюзивной реакции $h+N \to X+N$, проинтегрированное по переменной $M_X$ (missing mass) по всему полубесконечному промежутку, начиная с упругого порога

$$\frac{d\sigma_N}{d\mathbf{q}^2} = \int_{m_h^2}^{\infty} dM_X^2 \frac{d\sigma_N}{d\mathbf{q}^2 dM_X^2}. \qquad (32)$$

Радиус дейтрона определяется через электромагнитный (зарядовый) формфактор дейтрона $\rho(\mathbf{q}^2)$ с помощью соотношения

$$\frac{2}{R_d^2} = \int d\mathbf{q}^2 \rho(4\mathbf{q}^2). \qquad (33)$$

Кроме того, одночастичное инклюзивное сечение нормировано таким образом, что

$$\frac{d\sigma_N}{d\mathbf{q}^2 dM_X^2}(M_X^2 = m_h^2)|_{\mathbf{q}^2=0} \simeq \frac{\sigma_{hN}^{tot\,2}}{16\pi}. \qquad (34)$$

Обратим внимание на то, что формула Грибова (31) не разделяет вклады в дефект полного сечения рассеяния на дейтроне, проистекающие от процессов упругого и неупругого взаимодействий налетающего адрона с нуклонами дейтрона. Как упругие, так и неупругие процессы в рамках реджевской феноменологии, которая использовалась Грибовым, описывались с помощью одного и того же динамического механизма, определяемого полюсом Померанчука. Если предположение о том, что асимптотика упругих процессов при энергиях порядка 10÷100 GeV определяется полюсом Померанчука, еще можно было как-то обосновать, поскольку полные адронные сечения в этой области энергий почти постоянны и, следовательно, можно было сослаться на выводы Померанчука о том, что обменами в $t$-канале с квантовыми числами, отличными от вакуумных, можно пренебречь, то распространение этого предположения на неупругие процессы представлялось весьма сомнительным. И действительно, уже первые попытки применить формулу Грибова для описания экспериментальных данных показали, что дефект полного сечения рассеяния на дейтроне, вычисленный с помощью формулы Грибова, превышал экспериментально измеренную величину. Проведенный анализ существующих экспериментальных данных привел авторов работ [5] к выводу, что для описания неупругих процессов необходимо учитывать вклад не только от померонного обмена, но и вклады от обменов другими реджеонами. При этом выяснилось, что для некоторых комбинаций реджеонов вклады от трехреджеонных диаграмм в дефект полного сечения рассеяния на дейтроне имеют противоположный знак по сравнению с вкладом от трехпомеронной диаграммы. Такие вклады в дефект полного сечения рассеяния на дейтроне были названы антитеневыми эффектами. Очевидно, что антитеневые эффекты приводили к частичной компенсации дефекта полного сечения рассеяния на дейтроне и, тем самым, к улучшению согласия теории с экспериментом. Более тщательный анализ экспериментальных данных, однако, показал, что учет всех возможных трехреджеонных диаграмм не позволяет устранить расхождение теории и эксперимента. Вывод, к которому пришли авторы работы [7], заключается в том, что для описания динамики неупругих процессов необходимо учитывать более сложные, нежели трехреджеонные диаграммы.

В конечном итоге, проблема описания дефекта полного сечения рассеяния на дейтроне в рамках реджевской феноменологии остается до настоящего времени нерешенной. Основная трудность, с которой столкнулась здесь реджевская феноменология, связана с описанием процессов дифракционной диссоциации. Самая популярная реджевская модель супер-критического померона, которая давала более или менее удовлетворительную феноменологическую параметризацию для полных адронных сечений, оказалась абсолютно непригодной при описании экспериментальных данных для сечений процессов дифракционной диссоциации в $\bar{p}p$ столкновениях при высоких энергиях. Критический анализ модели супер-критического померона и дискуссию по данной проблематике можно посмотреть в нашей работе [8].

Совершенно ясно, однако, что вопросы о том, как устроен дефект полного сечения рассеяния на составных системах, каковыми являются ядра, какова структура возникающих при этом теневых поправок – это вопросы фундаментальной важности. Теоретическое понимание этих вопросов позволило бы нам, во-первых, создать правильное представление о структуре составных ядерных систем, а с другой стороны привело бы к более глубокому пониманию известных и обнаружению новых, ранее неизвестных фундаментальных динамических свойств нуклонов, являющихся структурными элементами составных ядерных систем.

## 3 Унитарность и трехчастичные силы в квантовой теории

В конце семидесятых годов прошлого века задача рассеяния на дейтроне была подробно рассмотрена нами с помощью динамических уравнений, которые возникают в результате редукции к единому времени формализма Бете-Солпитера для системы трех частиц в квантовой теории поля [9, 10, 11]. Первое и очень существенное обстоятельство, с которым мы столкнулись, было связано с тем, что последовательное рассмотрение проблемы трех тел в рамках релятивистской квантовой теории с необходимостью приводит к тому, что динамика трехчастичной системы должна обязательно включать кроме парных взаимодействий частиц в системе также новые фундаментальные трехчастичные силы, которые невозможно выразить через парные взаимодействия частиц. На этом пути было установлено, что фундаментальные трехчастичные силы связаны со специфическими неупругими взаимодействиями в двухчастичных подсистемах и определяют динамику специальных неупругих процессов взаимодействия двух частиц, известных под названием одночастичных инклюзивных реакций. При весьма общих предположениях нам удалось вычислить вклад трехчастичных сил в полное сечение рассеяния на дейтроне [11] и получить новую, чрезвычайно простую и изящную формулу для дефекта полного сечения рассеяния с ясной и прозрачной физической интерпретацией. Полученная структура теневых поправок к полному сечению рассеяния на дейтроне позволила обнаружить новые масштабные (скейлинговые) закономерности [12] фундаментального характера в процессах взаимодействия составных ядерных систем, речь о которых пойдет ниже. Кроме того, быть может одним из самых приятных для нас обстоятельств было и то, что уже предварительное сравнение с экспериментальными данными по полным протон-дейтронным и антипротон-дейтронным сечениям при высоких энергиях показало блестящее согласие теории с экспериментом [12, 13]. К этому обстоятельству мы также еще вернемся ниже.

Последовательность наших исследований с достаточно сложными, трудоемкими выкладками и утомительными расчетами можно проследить по цитированным выше работам [9, 10, 11, 12, 13]. Как это часто бывает, когда окончательный результат удается представить в виде очень простой формулы, в которую входят только экспериментально измеряемые величины с четкой и ясной физической интерпретацией, всегда возникает догадка о том, как этот результат можно было бы получить с помощью простых и наглядных рассуждений, не прибегая к решению сложных динамических уравнений с использованием математического аппарата квантовой теории поля. Прежде чем предъявить основные результаты наших исследований, мы хотели бы привести некоторые достаточно строгие рассуждения о том, какова могла бы быть самая общая структура для дефекта полного сечения рассеяния на дейтроне. Здесь может быть следовало бы еще раз напомнить, что Глаубером во введении к своей работе [1] также были представлены полуклассические рассуждения, обосновывающие результат сделанных им вычислений в рамках предложенной дифракционной теории.

Основную роль в наших рассуждениях будут играть фундаментальное требование квантовой теории, известное как унитарность $S$-матрицы или оператора рассеяния, и оптическая теорема, которая связывает полное сечение $\sigma_{AB}^{tot}$ взаимодействия двух систем $A$ и $B$ с мнимой частью амплитуды упругого рассеяния одной системы, скажем $A$, на другой системе $B$ на нулевой угол. В качестве систем $A$ и $B$ могут выступать произвольные адроны, например, пионы, каоны, нуклоны, либо сложные комплексы, составленные из адронов, например, ядра. Из соотношения унитарности, которому удовлетворяет амплитуда упругого рассеяния, следует, что полное сечение $\sigma_{AB}^{tot}$ представимо в виде суммы двух слагаемых

$$\sigma_{AB}^{tot} = \sigma_{AB}^{el} + \sigma_{AB}^{inel}, \qquad (35)$$

где $\sigma_{AB}^{el}$ есть полное сечение упругого взаимодействия, а $\sigma_{AB}^{inel}$ есть полное сечение неупругого взаимодействия систем $A$ и $B$. Величину $\sigma_{AB}^{inel}$ иногда называют полным сечением поглощения и используют обозначение $\sigma_{AB}^{abs}$. Представление (35) является однозначным и единственным. Для дефекта полного сечения рассеяния произвольного адрона $h$ на дейтроне, определенного формулой (7), тогда немедленно получаем, что он также представим в виде суммы двух слагаемых

$$\delta\sigma_{hd} = \delta\sigma_{hd}^{el} + \delta\sigma_{hd}^{inel}. \qquad (36)$$

Величину $\delta\sigma_{hd}^{el}$ мы будем называть упругим дефектом, а величину $\delta\sigma_{hd}^{el}$ – неупругим дефектом. Физическая интуиция подсказывает нам, что упругий дефект $\delta\sigma_{hd}^{el}$ должен определяться упругими взаимодействиями с нуклонами дейтрона и проистекать от процессов упругого перерассеяния налетающего адрона на нуклонах дейтрона, в то время как неупругий дефект $\delta\sigma_{hd}^{inel}$ обязан своим происхождением неупругим взаимодействиям с нуклонами дейтрона и должен быть связан с процессами дифракционной диссоциации налетающего адрона при столкновении с нуклонами дейтрона.

Попытаемся теперь понять, через какие физические величины можно было бы выразить упругий и неупругий дефекты. Опять же, если исходить из интуитивно ясной физической картины рассеяния на дейтроне, то очевидно, что упругий дефект должен быть пропорционален полному сечению упругого взаимодействия с нуклоном дейтрона. Коэффициент пропорциональности мы обозначим $a_{el}$, этот коэффициент будет определять вес, с которым полное сечение упругого взаимодействия с нуклоном будет входить в дефект полного сечения рассеяния на дейтроне. Кроме того, необходимо учесть комбинаторный фактор. В случае дейтрона этот фактор, очевидно, равен 2, поскольку налетающий адрон может упруго провзаимодействовать сначала с одним нуклоном, а затем с другим, либо то же самое в обратном порядке. Таким образом,

$$\delta\sigma_{hd}^{el} = 2\, a_{el}\, \sigma_{hN}^{el}. \tag{37}$$

Если данные рассуждения буквально применить к неупругому дефекту, то, казалось бы, мы должны были бы написать $\delta\sigma_{hd}^{inel} = 2\, a_{inel}\, \sigma_{hN}^{inel}$, однако это было бы неправильно. Дело в том, что среди всех возможных каналов неупругого взаимодействия налетающего адрона с нуклоном дейтрона мы должны учесть только те каналы, которые связаны с процессами дифракционной диссоциации, не повлекшими развал дейтрона. Кроме того, мы должны иметь в виду, что другие каналы неупругого взаимодействия налетающего адрона с нуклонами дейтрона уже включены в первые два слагаемых в формуле (7). В результате, правильное выражение для неупругого дефекта должно иметь вид

$$\delta\sigma_{hd}^{inel} = 2\, a_{inel}\, \sigma_{hN}^{sd}, \tag{38}$$

где $\sigma_{hN}^{sd}$ есть сечение дифракционной диссоциации на нуклоне. Коэффициент $a_{inel}$ определяет вес, с которым сечение дифракционной диссоциации входит в дефект полного сечения рассеяния на дейтроне.

Далее, что можно было бы сказать о коэффициентах $a_{el}$ и $a_{inel}$. Начнем с $a_{el}$. Во-первых, $a_{el}$, вообще говоря, может зависеть от энергии. Зависимость $a_{el}$ от энергии удобно выразить через единственную, зависящую от энергии характеристику упругого рассеяния, каковой является наклон дифракционного конуса $B_{el}$, простым образом связанный с эффективным радиусом упругого взаимодействия $B_{el} = R_{el}^2/2$. Поскольку $a_{el}$ является безразмерной величиной, то зависимость от $B_{el}$ необходимо чем-то обезразмерить. Единственной размерной величиной, которая позволяет это сделать в данной задаче и от которой также может зависеть $a_{el}$, так как налетающий адрон упруго взаимодействует не со свободным нуклоном, а с нуклоном, находящимся в ядре, – это радиус дейтрона. Итак, самая общая зависимость, которую можно было бы представить для $a_{el}$, выглядит следующим образом: $a_{el} = a_{el}(x_{el}^2)$, $x_{el}^2 = 2B_{el}/R_d^2 = R_{el}^2/R_d^2$. Аналогичным образом для $a_{inel}$ получаем $a_{inel} = a_{inel}(x_{inel}^2)$, $x_{inel}^2 = 2B_{sd}/R_d^2 = R_{inel}^2/R_d^2$, где $B_{sd}$ – наклон дифракционного конуса для процессов дифракционной диссоциации, связанный с эффективным радиусом неупругого взаимодействия соотношением $B_{sd} = R_{inel}^2/2$. В итоге, собирая формулы (36,37,38) вместе, для дефекта полного сечения рассеяния на дейтроне получаем

$$\delta\sigma_{hd} = 2\, a_{el}(x_{el}^2)\, \sigma_{hN}^{el} + 2\, a_{inel}(x_{inel}^2)\, \sigma_{hN}^{sd}, \tag{39}$$

$$x_{el}^2 = \frac{R_{el}^2}{R_d^2} = \frac{2B_{el}^2}{R_d^2}, \quad x_{inel}^2 = \frac{R_{inel}^2}{R_d^2} = \frac{2B_{sd}^2}{R_d^2}.$$

Здесь уместно сделать следующее замечание. В наших рассуждениях, конечно же, предполагалось, что как при упругих, так и при неупругих взаимодействиях налетающего адрона с нуклонами дейтрона, протон и нейтрон – динамически неразличимы, т.е. соответствующие динамические характеристики для протона и нейтрона являются одинаковыми, например, $\sigma_{hp}^{el} = \sigma_{hn}^{el} = \sigma_{hN}^{el}$, $B_{p,el} = B_{n,el} = B_{el}$ и т.д. Такое предположение является вполне оправданным, когда взаимодействия происходят при достаточно больших энергиях. Ясно, что при очень низких энергиях необходимо учитывать, что протон и нейтрон имеют разные массы и электрический заряд, однако формула (39) допускает естественную модификацию и на этот случай.

Функции $a_{el}$ и $a_{inel}$, стоящие в правой части формулы (39), естественно назвать упругими и, соответственно, неупругими структурными функциями дейтрона. Эти функции имеют ясный физический смысл. Функция $a_{el}$ является своего рода «счетчиком», который отсчитывает относительную долю событий, связанных с упругими перерассеяниями налетающего адрона на нуклонах дейтрона, среди всех возможных взаимодействий с дейтроном как целое. Эта функция зависит от переменной, которая является эффективным радиусом упругого взаимодействия с нуклоном, измеренным с помощью «линейки» с масштабом, определяемым радиусом дейтрона. При каждом значении этой переменной (при данном значении энергии) величина функции $a_{el}$ определяет вес, с которым полное сечение упругого взаимодействия с нуклоном при данной энергии входит в дефект полного сечения рассеяния на дейтроне. Такая же физическая интерпретация с очевидными изменениями в терминах переносится на неупругую структурную функцию $a_{inel}$. Функция $a_{inel}$ также представляет собой своего рода «измерительный прибор», но другой, который считает относительную долю других событий среди всех возможных взаимодействий с дейтроном, связанных с процессами неупругого взаимодействия с нуклонами дейтрона инклюзивного типа в области дифракционной диссоциации. Неупругая структурная функция зависит от другой скейлинговой переменной, которая является эффективным радиусом неупругого взаимодействия с нуклоном, измеренным с помощью «линейки» с тем же масштабом, определяемым радиусом дейтрона. Величина функции $a_{inel}$ при заданном значении энергии определяет вес, с которым сечение дифракционной диссоциации на нуклоне при той же энергии входит в дефект полного сечения рассеяния на дейтроне. Формула (39) может служить в качестве инструментария для экспериментального исследования структурных функций $a_{el}$ и $a_{inel}$ путем измерения дефекта полного сечения рассеяния на дейтроне с использованием экспериментальной информации об упругих сечениях рассеяния и сечениях дифракционной диссоциации на нуклоне. Для этих целей, однако, крайне важно было бы иметь надежное теоретическое обоснование этой формулы. Надо сказать, замечательно, что такое теоретическое обоснование действительно может быть получено.

Плод наших исследований, о которых шла речь выше, собственно, в том и состоял, что был найден способ строгого вывода формулы (39) в рамках квантовой теории поля. При этом вскрылось чрезвычайно любопытное обстоятельство, которое заключается в том, что неупругий дефект в полном сечении рассеяния на дейтроне является проявлением фундаментальных трехчастичных сил. Кроме того, в то же самое время было установлено, что трехчастичные силы определяют динамику одночастичных инклюзивных реакций. Мы приведем здесь полученную нами формулу, устанавливающую связь одночастичного инклюзивного сечения с амплитудой трехчастичных сил, которая имеет вид

$$\frac{2s}{\pi}\frac{d\sigma_{hN\to XN}}{dtdM_X^2} = \frac{(2\pi)^3}{I(s)}\Im\mathcal{F}_0^{scr}(\bar{s}; -\mathbf{q}, \mathbf{q}, \mathbf{p}; \mathbf{q}, -\mathbf{q}, \mathbf{p}), \tag{40}$$

$$\mathcal{F}_0^{scr} = -\Big[T_0 + \sum_{N=p,n}(T_0 G_0 T_{hN} + T_{hN} G_0 T_0)\Big]_{on\,energy\,shell},$$

где $\mathbf{p}$ есть импульс налетающего адрона $h$, $s = (p+q_N) \simeq 2m_N|\mathbf{p}|$, $t = -4\mathbf{q}^2$, $\bar{s} = 2(s+m_N^2) - M_X^2$, $I(s) = 2\lambda^{1/2}(s, m_h^2, m_N^2)$, $T_{hN}$ и $T_0$ – амплитуды упругого рассеяния на нуклоне и амплитуда трехчастичных сил в системе трех частиц, в которую входят адрон $h$ и два нуклона, вне поверхности энергии. Одночастичное инклюзивное сечение в нашем случае определено таким образом, что оно удовлетворяет следующему условию нормировки

$$\int dt dM_X^2 \frac{d\sigma_{hN\to XN}}{dt dM_X^2} = <n_N> \sigma_{hN}^{inel}, \tag{41}$$

где $<n_N>$ – средняя множественность нуклонов, рождающихся в $hN$ столкновении, а интегрирование в левой части формулы (41) проводится по всей кинематически разрешенной области. Сечение дифракционной диссоциации, как известно, определяется путем обрезания интервала интегрирования по переменной $M_X^2$

$$\sigma_{hN}^{sd}(s) = \int_{M_{min}^2}^{\varepsilon s} dM_X^2 \int_{t_-(M_X^2)}^{t_+(M_X^2)} dt \frac{d\sigma_{hN\to XN}}{dt dM_X^2}, \tag{42}$$

где величина параметра $\varepsilon$, определяющего верхний предел интегрирования по переменной $M_X^2$, варьируется в разных работах в пределах 0.05÷0.15, нет общего соглашения относительно выбора величины этого параметра. Однако в рамках используемого нами формализма условие согласованности диктует строго определенное значение для этого параметра, которое определяется из равенства

$$\varepsilon = \varepsilon^{ex} = \sqrt{2\pi}/2M_N R_d \simeq 0.125. \tag{43}$$

Экспериментально установлено, что одночастичные инклюзивные сечения в области дифракционной диссоциации очень хорошо параметризуются с помощью формулы

$$\frac{2s}{\pi}\frac{d\sigma}{dt dM_X^2} = A(s, M_X^2)\exp[b(s, M_X^2)t]. \tag{44}$$

С помощью формулы (40) можно выразить величины $A(s, M_X^2)$ и $b(s, M_X^2)$ через основные характеристики фундаментальных трехчастичных сил. В частности, для наклона дифракционного конуса $b(s, M_X^2)$ таким образом получаем

$$b(s, M_X^2) = \frac{R_0^2(\bar{s})}{2}, \quad \bar{s} = 2(s+m_N^2) - M_X^2, \tag{45}$$

где $R_0$ – эффективный радиус трехчастичных сил. Заметим, что точно таким же соотношением, как (45), наклон дифракционного конуса в упругом рассеянии связан с эффективным радиусом парных сил.

Использованный нами формализм позволяет провести аналитические вычисления до конца, если для этих целей воспользоваться надежно установленными на эксперименте параметризациями (26,44). На этом пути нам удалось получить чрезвычайно простые формулы для структурных функций $a_{el}$ и $a_{inel}$, которые имеют вид

$$a_{el}(x^2) = \frac{x^2}{1+x^2}, \quad a_{inel}(x^2) = \frac{x^2}{(1+x^2)^{\frac{3}{2}}}. \tag{46}$$

Хотя явный аналитический вид (46) для $a_{el}$ и $a_{inel}$ был получен в предположении, что амплитуды $T_0$ и $T_{hN}$ при больших энергиях, будучи суженными на поверхность энергии, становятся чисто мнимыми, учет реальных частей этих амплитуд не представляет туда и соответствующее обобщение также было получено [13].

## 4 Анализ и обсуждение структурных функций, сравнение с экспериментом

Результат вычислений функций (46) и анализ этих функций, однако, достойны отдельного обсуждения. Во-первых, возвращаясь к формуле (39), заметим, что формула Глаубера получается из нее, если в этой формуле пренебречь неупругим дефектом, а для упругой структурной функции воспользоваться приближением $a_{el}(x^2) \simeq x^2$, справедливым при $x^2 << 1$, и учесть, что $\sigma_{hN}^{el} \simeq \sigma_{hN}^{tot\,2}/16\pi B_{el}$. Во-вторых, необходимо обратить внимание на то, что структурные функции $a_{el}$ и $a_{inel}$ имеют совершенно разное поведение: $a_{el}(x^2)$ является монотонной (возрастающей) функцией при изменении аргумента на полубесконечном интервале $0 \leq x^2 < \infty$, а область ее значений ограничена интервалом $0 \leq a^{el} \leq 1$, в то время как функция $a_{inel}$ сначала возрастает, достигает максимума при $x^2 = 2$, а затем убывает, исчезая на бесконечности, при этом область ее значений лежит в интервале $0 \leq a^{inel} \leq 2/3\sqrt{3}$. Разумеется, что такое различие в поведении структурных функций $a_{el}$ и $a_{inel}$ приводит к далеко идущим физическим следствиям. Например, при сверхвысоких энергиях, соответствующих $x^2 \to \infty$, мы находим, что неупругий дефект исчезает, величина упругого дефекта стремится к удвоенному значению полного упругого сечения рассеяния на нуклоне, а полное сечение рассеяния на дейтроне приближается к удвоенному значению полного сечения поглощения на нуклоне. Следовательно, при сверхвысоких энергиях должна восстанавливаться $A$ зависимость полных сечений рассеяния на ядрах с той лишь разницей, что фундаментальной величиной, стоящей при $A$, является не полное сечение рассеяния на нуклоне, а полное сечение поглощения на нуклоне:

$$\sigma_{hA}^{tot} = A\sigma_{hN}^{inel}, \, s \to \infty. \tag{47}$$

Конечно, вне всякого сомнения, сравнение полученных нами теоретических результатов с имеющимися экспериментальными данными по полным сечениям рассеяния протонов и антипротонов на дейтронах представляло для нас особый интерес. На рисунках 1 и 2 представлены предварительные результаты такого сравнения. Кривые на этих рисунках соответствуют полным сечениям рассеяния протонов и антипротонов на дейтронах, вычисленным по формулам (7, 39, 46). При этом мы использовали сделанное нами ранее [8] глобальное описание полных $pp$ и $\bar{p}p$ сечений и сечений дифракционной диссоциации с учетом самых последних экспериментальных данных, полученных коллаборацией CDF во ФНАЛе [14]. Кроме того, по данному поводу, наверное, следовало бы еще добавить, что сравнение с экспериментальными данными по полным сечениям рассеяния протонов и антипротонов на дейтронах проводилось нами, как бы, в два этапа. На первом этапе теоретические расчеты сравнивались с экспериментальными данными по полным сечениям рассеяния только антипротонов на дейтронах в предположении, что $R_d^2$ является единственным свободным параметром, величина которого должна была определяться из подгонки экспериментальных данных. В результате статистического анализа для величины $R_d^2$ было получено следующее значение: $R_d^2 = 66.61 \pm 1.16\,\text{GeV}^{-2}$. Здесь уместно обратить внимание на следующее обстоятельство. Последние экспериментальные измерения радиуса материи дейтрона свидетельствуют: $r_{d,m} = 1.963(4)\,\text{fm}$ [15], откуда следует, что $r_{d,m}^2 = 3.853\,\text{fm}^2 = 98.96\,\text{GeV}^{-2}$. Полученное нами значение для $R_d^2$ удовлетворяет соотношению $R_d^2 = 2/3\,r_{d,m}^2$. Для полноты, результаты теоретических расчетов представлены на рисунке 1 вплоть до энергий Теватрона ФНАЛ. На втором этапе экспериментальные данные по полным сечениям рассеяния протонов на дейтронах сравнивались с теоретическими расчетами, в которых значение величины $R_d^2$ было зафиксировано на том численном значении, которое было получено на первом этапе из анализа данных по полным $\bar{p}d$ сечениям. Другими словами, кривая на рисунке 2 соответствует теоретическим расчетам, сделанным с помощью формул (7, 39, 46), в которых не было ни одного свободного параметра. На этом рисунке результаты теоретических расчетов также представлены вплоть до энергий Теватрона ФНАЛ. Как видно, рисунки 1 и 2 свидетельствуют о блестящем согласии теории и эксперимента.

На рисунке 3 представлены результаты выполненных нами теоретических расчетов упругого и неупругого дефектов полного сечения рассеяния (анти)протонов на дейтронах в интервале энергий $\sqrt{s} \sim 10 \div 2000$ GeV. Из этих расчетов следует, что величина упругого дефекта составляет около 10% от величины полного нуклон-нуклонного сечения, а величина неупругого дефекта составляет около 10% от величины упругого дефекта, т.е. примерно 1% от величины полного нуклон-нуклонного сечения. Образно выражаясь, можно было бы сказать, что если упругий дефект представляет собой тонкую структуру, то неупругий дефект следовало бы отнести к сверхтонкой структуре в полных сечениях рассеяния на дейтроне. В нашем подходе неупругий дефект связан с проявлением фундаментальных трехчастичных сил, поэтому в этом смысле трехчастичные силы играют роль «тонкой подстройки» в динамике релятивистской трехчастичной системы. Надо отдать должное физикам-экспериментаторам, создающим экспериментальные установки, в которых достигается точность измерений, позволяющая дискриминировать неупругие дефекты в полных сечениях рассеяния частиц при высоких энергиях. В этой связи, дальнейшие экспериментальные, высокоточные измерения адрон-дейтронных полных сечений при высоких энергиях представляется нам чрезвычайно важными. Было бы также крайне желательным создание ускоренных пучков дейтронов на действующих в настоящее время и строящихся ускорителях и коллайдерах.

Как уже отмечалось нами выше, максимальное значение неупругого дефекта достигается при $x_{inel}^2 = 2$ ($x_{inel}^2 \equiv R_0^2/R_d^2$). Иначе говоря, значение энергии, при которой неупругий дефект принимает максимальное значение, определяется из уравнения $R_0^2(s_{max}) = 2R_d^2$. Вычисления, сделанные с учетом проведенного нами анализа существующих экспериментальных данных, дают $\sqrt{s_{max}} = 9.01\,10^8\,\text{GeV} = 901\,\text{PeV}$. Очевидно, что такие значения энергий не достижимы на ныне действующих и проектных ускорителях. Однако проявления данного эффекта можно поискать в явлениях, которые наблюдаются в космических лучах сверхвысоких энергий, но это тема для отдельных исследований, которые в настоящее время проводятся. Заметим, однако, что величина $s_{max}$ имеет ясный физический смысл, она разделяет две области по энергии: область энергий $s < s_{max}$, при которых эффективный радиус трехчастичных сил не превышает размеров дейтрона, точнее $1/2\,R_0^2(s) < R_d^2$, и область энергий $s > s_{max}$, при которых эффективный радиус трехчастичных сил становится больше размеров дейтрона $1/2\,R_0^2(s) > R_d^2$. Физический механизм, который приводит к убыванию неупругого дефекта при $s > s_{max}$, связан с экранировкой трехчастичных сил парными силами, которая возникает в результате двухчастичного взаимодействия в начальном и конечном состояниях трехчастичной системы до и после трехчастичного взаимодействия. Особо подчеркнем, что данный механизм однозначно диктуется требованием унитарности. Требование унитарности, таким образом, приводит к подавлению неупругого дефекта при сверхвысоких энергиях так, что от полного дефекта при асимптотически бесконечных энергиях остается только упругая часть. Существование границы $s_{max}$, начиная с которой происходит подавление неупругого дефекта, представляется нам чрезвычайно важной характеристикой фундаментальной динамики. На рисунке 4 представлен теоретически вычисленный нами неупругий дефект в области максимума.

Из формулы Глаубера (8) следует, что с уменьшением межнуклонного расстояния в дейтроне величина упругого дефекта растет. Но именно конфигурации с малыми межнуклонными расстояниями в дейтроне наиболее благоприятны для проявления собственно трехчастичного взаимодействия. Когда эффективный радиус взаимодействия налетающего адрона с нуклоном становится сравнимым с межнуклонным расстоянием, картина упругих перерассеяний на нуклонах дейтрона перестает быть адекватной полной картине взаимодействия с дейтроном. В этом случае необходимо также учитывать собственно трехчастичные силы. Очевидно, что в дейтроне динамически возможны конфигурации, когда нуклоны находятся вблизи друг от друга, но теория Глаубера не позволяет учитывать такие конфигурации. Мы уже говорили, что рассмотрение и учет таких конфигураций требует более детального исследования динамики процессов рассеяния на дейтроне. Аппарат динамических уравнений в квантовой теории поля, который мы использовали, как раз позволяет такие детальные исследования провести. Еще раз подчеркнем, что важная роль в наших исследованиях отводилась концептуальному понятию фундаментальных трехчастичных сил, которые с необходимостью возникают при последовательном рассмотрении динамики системы трех частиц в рамках релятивистской квантовой теории. Установленная при этом связь фундаментальных трехчастичных сил с динамикой одночастичных инклюзивных реакций сама по себе представляет важный результат, полученный, как бы, мимоходом. Этот результат тем более важен, что может служить основой как для создания методов аналитических вычислений, так и для разного рода феноменологических исследований.

Проведенное нами сравнение теории с экспериментальными данными по полным нуклон-дейтронным сечениям показывает, что для описания рассеяния частиц на дейтроне при высоких энергиях достаточно учитывать только нуклонные степени свободы в дейтроне. Слабосвязанная двухнуклонная система дейтрон выглядит так, что кластеризация кварков в нуклоны не нарушается даже тогда, когда нуклоны подходят близко друг к другу. Нуклоны, находясь вблизи друг от друга в дейтроне, не теряют своей индивиду-альности и поэтому не возникает необходимость вводить обезличенные шестикварковые конфигурации в дейтроне. Полученная нами структура для дефекта полного сечения рассеяния на дейтроне отвечает именно такой картине. Мы показали, что общий формализм квантовой теории поля допускает возможность представления динамики рассеяния частицы на составной системе через фундаментальную динамику рассеяния частицы на изолированных конституентах и структуру самой составной системы. Хотя детально была рассмотрена динамика рассеяния частицы на двухчастичной составной системе дейтроне, использованный нами общий формализм допускает естественное обобщение на более сложные многочастичные, составные, ядерные системы. Конечно, сложность рассмотрения при этом существенно возрастает.

## 5 Заключительные замечания

Здесь, по-видимому, было бы уместно напомнить, что ранее неоднократно предпринимались попытки понять статус теории Глаубера в рамках строгой нерелятивистской теории системы многих частиц, в которой полная амплитуда представляется в виде бесконечного итерационного ряда (ряда Ватсона) многократных рассеяний. Говоря проще, вопрос, который необходимо было выяснить, заключался в следующем: при каких допущениях, предположениях и приближениях ряд многократных рассеяний можно было бы свести к глауберовской форме для амплитуды рассеяния. Одно, но очень важное приближение, которое необходимо было сделать, напрашивалось самой постановкой вопроса и аналогией с оптикой, лежащей в основе теории Глаубера. Функцию Грина налетающей частицы, которая представляется Фурье образом от пропагатора частицы

$$G_0(E|\mathbf{r};\mathbf{r}') = \frac{1}{(2\pi)^3}\int d\mathbf{p}\frac{e^{i\mathbf{p}(\mathbf{r}-\mathbf{r}')}}{p^2 - q^2 - i0} = \frac{1}{4\pi}\frac{e^{iq|\mathbf{r}-\mathbf{r}'|}}{|\mathbf{r}-\mathbf{r}'|}, \quad q^2 = 2mE, \tag{48}$$

было бы естественно заменить более простым выражением, учитывающим движение частицы лишь вдоль прямолинейной траектории. Такое выражение получается, если в пропагаторе частицы сделать приближение

$$\frac{1}{p^2 - q^2 - i0} = \frac{1}{\Delta^2 + 2\mathbf{q}\mathbf{\Delta} - i0} \simeq \frac{1}{2\mathbf{q}\mathbf{\Delta} - i0}, \quad \mathbf{\Delta} \equiv \mathbf{p} - \mathbf{q}, \tag{49}$$

которое является обоснованным при выполнении условия $\Delta/2q << 1$. В этом случае вычисление интеграла в формуле (48) приводит к выражению

$$G_0(E|\mathbf{r};\mathbf{r}') \simeq \frac{i}{2q}\delta^2(\mathbf{r}_\perp - \mathbf{r}'_\perp)\theta(z-z')e^{iq_z(z-z')}. \tag{50}$$

Замечательная особенность приближенного выражения (50) для функции Грина состоит в том, с его помощью удается найти точное решение уравнения Липпмана-Швингера для волновой функции в теории потенциального рассеяния. При этом для амплитуды потенциального рассеяния естественным образом возникает известное из оптики эйкональное представление. Кроме того, было установлено, что использование приближенной функции Грина (50) в рядах многократного рассеяния для амплитуд рассеяния на ядрах приводит к тому, что все эффекты, связанные с выходом распространяющейся в ядре частицы за поверхность энергии пропадают и остаются только амплитуды рассеяния на нуклонах на поверхности энергии. Здесь же необходимо упомянуть важный результат Харрингтона [16], который показал, что в случае рассеяния на дейтроне в разложении Ватсона внеэнергетический вклад от члена с двухкратным перерассеянием в дифракционном пределе в точности сокращается с остатком бесконечного ряда многократных перерассеяний, начиная с трехкратного перерассеяния. Этот результат наводит нас на мысль, что формула (40) для одночастичного инклюзивного сечения, полученная нами как приближенная, на самом деле может оказаться точной.

Идея использовать приближенную функцию Грина (50) вместо точной (48) нашла свое применение также в квантовой теории поля при построении эффективных методов суммирования бесконечных рядов амплитуд, соответствующих широкому классу фейнмановских диаграмм [17, 18, 19, 20]. В работе [21] можно найти строгий математический вывод в квантовой теории поля эйконального представления для амплитуды упругого рассеяния двух частиц при высоких энергиях.

И последнее замечание, которое нам хотелось бы сделать, связано с тем, что в литературе по обсуждаемой здесь тематике не редко можно встретить бездоказательные утверждения, скорее даже спорные мнения. Например, предположение о том, что теневые поправки в случае дейтрона должны исчезать в пределе бесконечных энергий так, что $\sigma_d^{tot} \to 2\sigma_N^{tot}$, $s \to \infty$, можно встретить не только в работе [2] (см. также [4]). Строгий результат, который был получен в наших исследованиях (о нем шла речь выше) заключается в том, что упругий дефект полного сечения рассеяния на дейтроне не только не исчезает, но только он и выживает при сверхвысоких энергиях, превышающих границу, определяемую максимальным значением неупругого дефекта. Точный результат состоит в том, что $\sigma_d^{tot} \to 2\sigma_N^{inel}$, $s \to \infty$.

В заключение, возвращаясь к началу нашего повествования, следовало бы сказать, что решение представить данный миниобзор было продиктовано, в том числе, нашим желанием выразить свое удовольствие признанием также выдающегося вклада лауреата Нобелевской премии по физике за 2005г. Роя Глаубера в создание и развитие дифракционной теории фундаментальных ядерных процессов при высоких энергиях. Свою точку зрения на явления дифракции мы уже высказывали [13]. Она состоит в том, что идеи и методы дифракционной теории доставляют уникальный и в то же время универсальный инструмент теоретических исследований, с помощью которого можно понять и осознать иерархию масштабов космоса и микрокосмоса, услышать новые звуки «музыки сфер» в том же смысле, о котором рассуждали древние пифагорейцы [22], и укрепить нашу веру в единство физической картины Мира.

# Список литературы

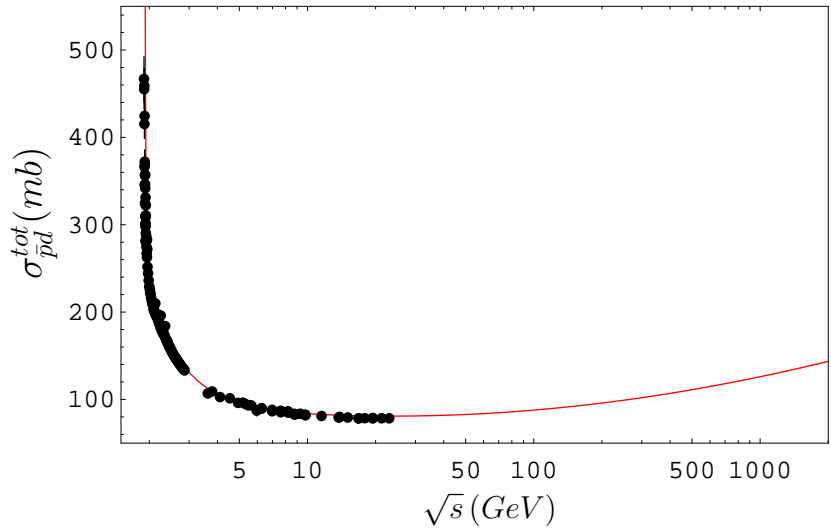

Рис. 1: Экспериментальные данные по полным сечениям рассеяния антипротонов на дейтронах. Кривая представляет результат сравнения теории с экспериментом.

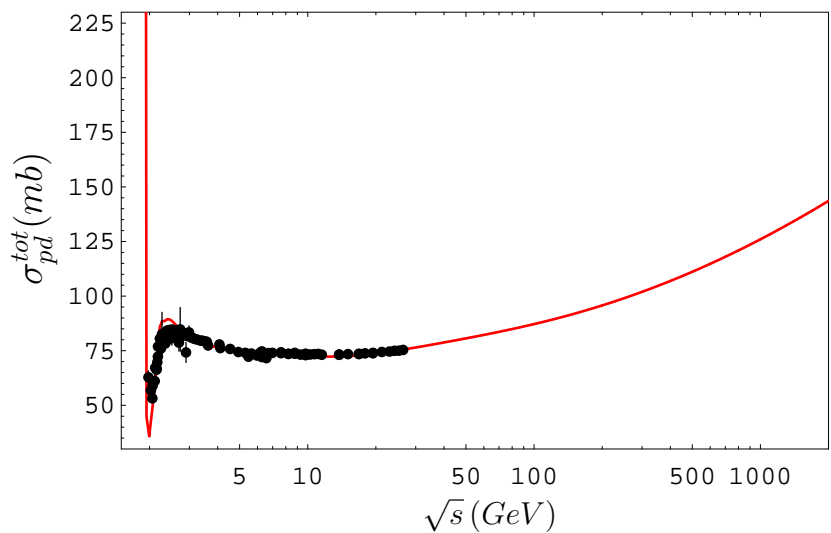

Рис. 2: Экспериментальные данные по полным сечениям рассеяния протонов на дейтронах. Кривая представляет результат сравнения теории с экспериментом.

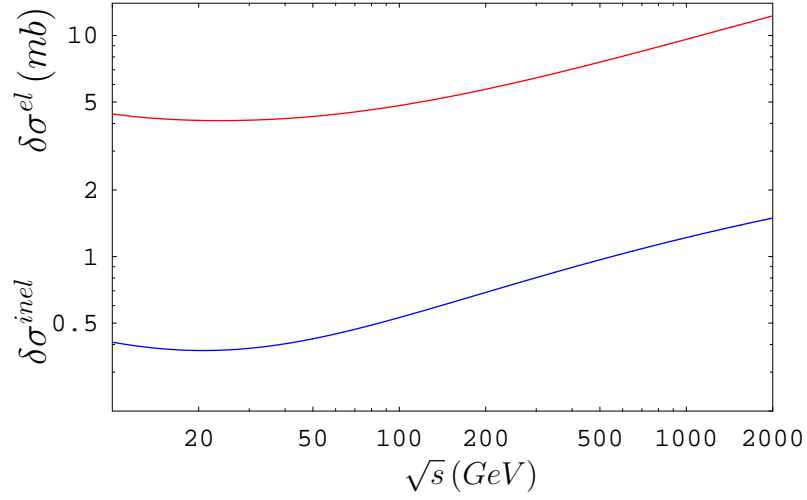

Рис. 3: Теоретически вычисленные упругий и неупругий дефекты полного сечения расеяния (анти)протонов на дейтроне.

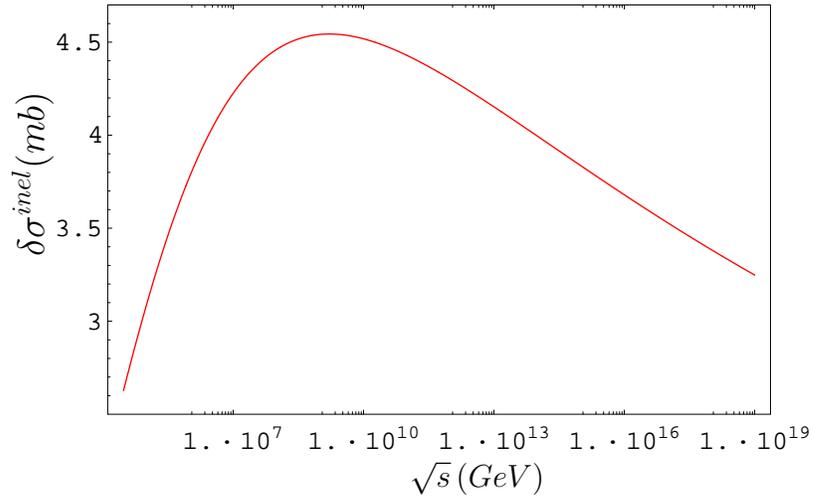

Рис. 4: Теоретически вычисленный неупругий дефект полного сечения расеяния (анти)протонов на дейтроне в области максимума.

22